# Two-photon absorption coefficient determination with a differential F-scan technique

**E Rueda,[1] J H Serna,[2] A Hamad and H Garcia[3,*]**

[1]*Grupo de Óptica y Fotónica, Instituto de Física, U de A, Calle 70 No. 52-21, Medellín, Colombia*
[2]*Grupo de Óptica y Espectroscopía, Centro de Ciencia Básica, Universidad Pontificia Bolivariana, Ca. 1 No. 70-01, Campus Laureles, Medellín, Colombia*
[3]*Department of Physics, Southern Illinois University, Edwardsville, Illinois, 60026, USA*
*hgarcia@siue.edu*

**Abstract:** In this paper we present a modification to the recently proposed transmission F-scan technique, the differential F-scan technique. In differential F-scan technique the programmed focal distance in the electronic-tunable lens oscillates, allowing the light detector of the setup to record a signal proportional to the derivative of the signal recorded with an F-scan. As for the differential Z-scan a background-free signal is obtained, but also the optical setup is simplified and the available laser power is double. We also present and validate a new fitting-procedure protocol that increments the accuracy of the technique. Finally, we show that fitting a signal from differential F-scan or the derivate of the signal of transmission F-scan is more accurate than simply fitting the signal from F-scan directly. Results from two-photon absorption at 790 nm of CdS, ZeSe and CdSe are presented.

## 1. Introduction

In the past decades, several optical techniques have been proposed to measure nonlinear optical properties such as two-photon absorption (TPA) coefficient for different types of materials, especially metals, organic and inorganic semiconductors [1,2]. Among them, Z-scan is a particularly widely used technique due to its relatively simple optical setup and data treatment [3]. It is based on the scanning of spatial beam modifications suffered by a laser beam after interacting with a sample, while it is focused and defocused in time: when the sample is near the focal point of the beam the high intensities generated produce nonlinear phenomena such variations in the refractive index and multi-photon absorption. However, some problems, related to laser fluctuations, beam alignment and mechanical vibrations can influence the results obtained, compromising the sensitivity of the technique. To overcome these limitations, modifications to the basic Z-scan setup were proposed in the following years. To enhance the sensitivity of the technique Xia et al. [4] replaced the far-field aperture in the standard Z-scan by an obscuration disk that blocks most of the beam, while Zhao et al. [5] used a top-hat beam instead of a Gaussian beam, and Martinelli et al. [6] measure the reflected beam from the sample in a Brewster angle configuration. To improve the signal to noise ratio some authors have used balance-detection systems [7,8], while Ménard et al. [9] introduce the differential Z-scan where a piezo-transducer device generates an oscillatory motion to induces a periodic modulation of the beam intensity at the sample, which in turn produces a modulation of the transmitted light proportional to the spatial derivative of the transmitted light, and therefore provides a background-free measurement. More recently, a new technique, that is a variation of Z-scan, was proposed. This method, which is called F-scan [10,11], use an electrically focus-tunable lens (EFTL) instead of a fixed lens to generate different focal points, allowing to replace the translation stage and leaving the sample fixed in space, i.e. eliminating mechanical movements from the setup. The focal distance in the EFTL is a function of the applied current to the lens. Analogously to Ménard et al. for the case of Z-scan, we present in this paper an open aperture differential F-scan technique (DF-scan) to determine the nonlinear absorption optical properties of materials. The setup has been oversimplified by using an EFTL and at the same time modulating the focal length of the EFTL with a rectangular low frequency signal that can be detected with a PSD (phase sensitive detector), in our case a lock-in amplifier, increasing the sensitivity of the system and reducing or eliminating laser fluctuations.

## 2. Transmission F-scan (TF-scan)

The F-scan experimental setup depicted in Fig. 1. is used for the determination of the two-photon absorption (TPA) coefficient (open aperture architecture).

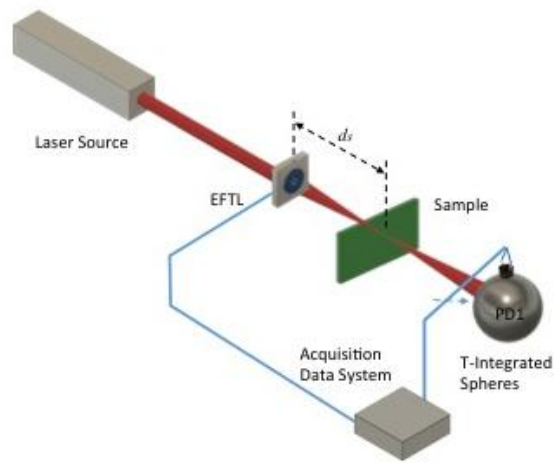

Fig. 1. TF-scan experimental optical setup for determination of TPA coefficients (open-aperture architecture).

A laser Gaussian beam modulated with a chopper impinges on an EFTL which is a lens that has the capability to vary its focal distance $f$ over a specific range when an electric current is applied to it (see Fig. 2), focusing the Gaussian beam at different positions. The sample is placed at a fixed position $d_s$ inside the range of the EFTL. The light transmitted through the sample is collected by a photodetector PD1. The output signal is then filtered with a Lock-in amplifier and processed with an acquisition data system.

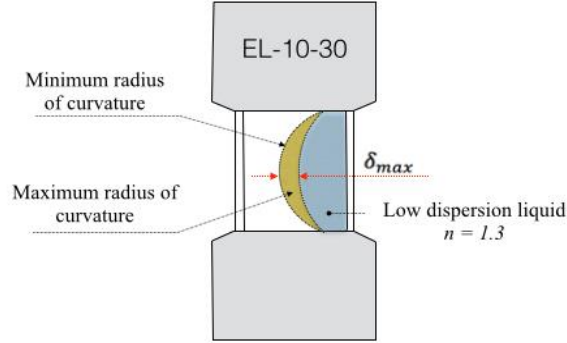

Fig. 2. OPTOTUNE-10-30-C electrically focus-tunable lens. This type of lens changes its shape (curvature) due to an optical fluid sealed off by a polymer membrane, when a current is applied.

To determine the TPA coefficient $\beta$ we measured the transmittance of the nonlinear medium as a function of the focal length, $f$, (see Fig. 1), collecting all light transmitted through the sample. When the distance $|d_s - f|$ is large the normalized transmittance has a value close to unity because linear optical effects are produced in the sample. In contrast, small values of $|d_s - f|$ imply that the laser beam is focused near the sample, thus increasing the optical intensity and generating nonlinear optical phenomena such as TPA. The TPA coefficient is obtained by fitting a theoretical curve (Eq. (9)) to the experimental data, using $\beta$ as the fitting parameter, and under the assumption that all the experimental parameters are known. A typical experimental curve is shown in Fig. 10(left).

### EFTL characterization

Fig. 3 shows the dependence of the EFTL optical power $\wp$ on the applied current and expose the disagreement between the experimental data and the data reported by the manufacturer. In the T-Fscan technique it is crucial to know the focal length with high accuracy in order to obtain correct values of $\beta$. Therefore, the characterization of the EFTL optical power on the applied current has to be done with high-precision.

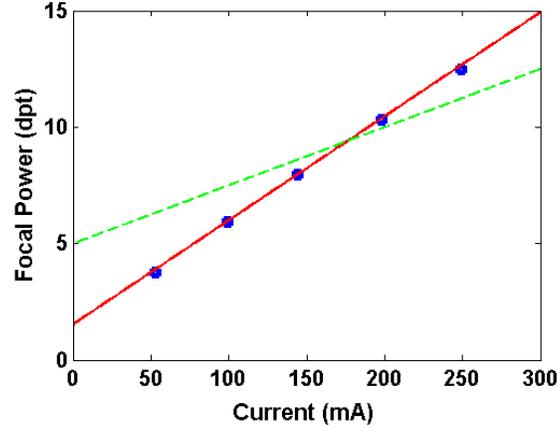

Fig. 3. The optical power, $\wp$, of the EFTL as a function of the applied current. (circles) Experimental data; (continuous line) fit of the experimental data using Eq. (1); (dashed line) data provided by Optotune Inc.

Different techniques can be used to obtain the correct dependence of the EFTL optical power on the applied current (circles in Fig. 3) by measuring the focal length $f$ as a function of current. We used a laser beam profiler to measure $f$ as a function of the applied current. Another way is to use the TPA phenomena: by placing a sample with a nonzero TPA coefficient at two different distances from the EFTL and finding the values of the applied current that produce the lowest intensity for each of the locations. The sample distance from the EFTL will correspond to the EFTL focal length that produced a minimum transmission. Then, using the relation $\wp = 1/f$, the EFTL optical power is obtained as a function of current. In our case the experimental data was fitted obtaining the following expression (continuous line in Fig. 3):

$$\wp = 0.045 J + 1.522 \qquad (1)$$

where $J$ is the applied current measured in mA. Another important experimental parameter for the correct determination of the nonlinear optical parameters is the beam-waist radius $w_0$. Typically, for spherical lenses and assuming that the beam has a spatial Gaussian profile, the radius of the beam at the beam waist is determined as a function of the focal length with the equation:

$$w_0(f) = \frac{2\lambda f}{\pi D} \qquad (2)$$

where $\lambda$ is the wavelength of the incident beam with spot diameter $D$. But being aware of the existence of optical aberrations on the optical system that distort the wavefront of the beam, a correction factor $C_f$ is needed in order to correctly calculate the beam waist. Thus, Eq. (2) is replaced by

$$w_0(f) = \frac{2\lambda f}{\pi D} C_f \qquad (3)$$

To determine the correction factor we used a laser beam profiler to measure the beam waist at each focal plane. Then, by using $C_f$ as the fitting parameter between the experimental data and Eq. (3), a correction factor $C_f = 1.36$ was obtained for the special case of our EFTL. Fig. 4 shows the difference between the corrected and non-corrected beam-waist diameter value as a function of the EFTL focallength.

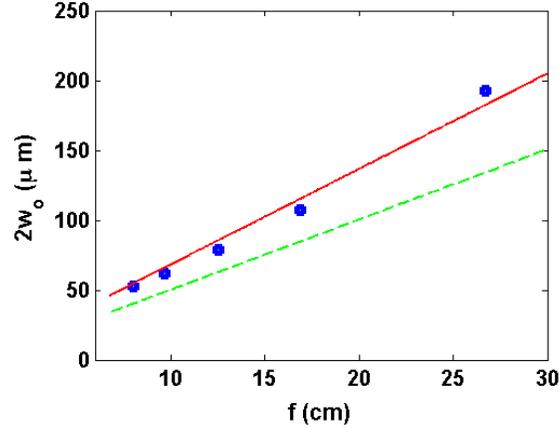

Fig. 4. Beam-waist diameter as a function of EFTL focal length. (circles) Experimental data measured with a laser beam profiler; (dashed line) beam waist diameter calculated with Eq. (2); (continuous line) beam waist diameter calculated with Eq. (3) and $C_f = 1.36$.

Once the beam-waist radius is correctly determined, it is possible to calculate with precision the beam radius $w(f)$ (Eq. (4)) at the sample surface for every programed EFTL focal length,

$$w(f) = \sqrt{1 + \left(\frac{d_s - f}{z_0(f)}\right)^2},  \quad (4)$$

where $z_0(f) = \pi w_0^2(f)/\lambda$ is the Rayleigh range. Fig. 5 shows the dependence of the beam radius at the sample location as a function of the EFTL focal length. Notice the difference between the results obtained with and without the corrected beam-waist radius. Also notice the asymmetry relative to the sample location. For $f$ values smaller than $d_s$ the radius at the sample increases faster than those for $f$ values larger than $d_s$. This will cause the shape of the TF-scan to be asymmetric around $d_s$. Therefore, experimentally, we must normalize the transmitted intensity relative to the intensity corresponding to the shortest focal lengths used in the experiment.

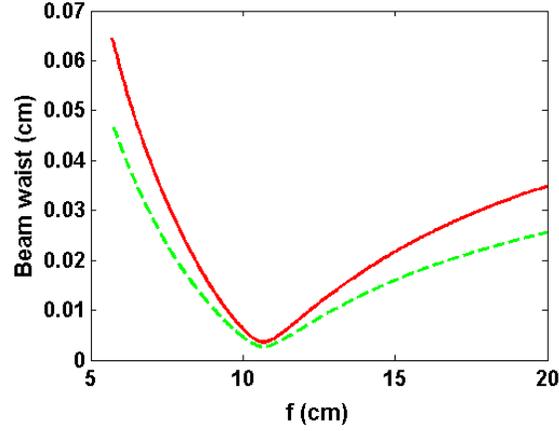

Fig. 5. Beam radius at the sample as a function of the EFTL focal length. The continuous and dashed lines are the corrected (Eq. (3)) and not corrected (Eq. (2)) beam waist radius $w_0$ respectively. For this plot we used $D = 2.0$ mm and $d_s = 10.7$ cm.

*Theoretical background*

The laser beam in our experimental setup has a Gaussian spatial profile and a hyperbolic secant temporal profile. Therefore, at the front surface of the sample, the incident beam intensity as a function of the EFTL focal length $f$ is given by:

$$I_{in}(r,f,t) = I_0(f)\exp\left[-2\left(\frac{r}{w(f)}\right)^2\right]\operatorname{sech}^2\left(\frac{t}{\tau_0}\right) \quad (5)$$

In the above equation $r$ is the radial position with respect to the optical axis, $t$ is time, $\tau_0 = \tau/(2\ln(1+\sqrt{2}))$, where $\tau$ is the full width at half-maximum pulse duration and $I_0(f)$ is the peak intensity of the beam at sample position as a function of the EFTL focal length:

$$I_0(f) = \frac{2\ln(1+\sqrt{2})P_{avg}}{\tau\nu\pi w^2(f)}. \quad (6)$$

Here $P_{avg}$ is the average power of the incident laser beam at the sample, and $\nu$ is the laser pulse repetition rate. The intensity at the exit surface of the sample can be written as:

$$I_{out}(r,f,t) = \frac{(1-R)^2 I_{in}(r,f,t)e^{-\alpha L}}{1+\beta(1-R)I_{in}(r,f,t)L_{eff}}, \quad (7)$$

where $L$ is the thickness of the sample, $R$ is the reflection coefficient of the sample, $\alpha$ is the linear absorption coefficient, $\beta$ is the two-photon absorption coefficient (TPA), and $L_{eff} = \left(1-e^{-\alpha L}\right)/\alpha$ is the effective sample thickness. Thus, the transmittance at the detector plane can be express as:

$$T(f) = \frac{1}{B(f)}\int_0^\infty \ln\left[1+B(f)\operatorname{sech}^2(\rho)\right]d\rho, \quad (8)$$

where $B(f) = \beta(1-R)I_0(f)L_{eff}$ and $\rho = 2\ln(1+\sqrt{2})t/\tau$. The transmittance given by Eq. (8) can be simplified when $B(f) < 1$ [12]:

$$T(f) = \sum_{m=0}^{N}\left\{\frac{[-B(f)]^m}{m+1}\prod_{n=0}^{m}\left[\frac{2(m-n)+\delta_{mn}}{2(m-n)+1}\right]\right\}. \quad (9)$$

As $B(f)$ gets closer to one, more terms of the sum are need it. Thus, one must use at least $N = 11$ in order to guarantee that the obtained value of $\beta$ is not underestimated. Otherwise, the obtained value of $\beta$ will be smaller than its actual value.

### 3. Differential F-scan (DF-scan)

To reduce noise in TF-scan, or in any intensity scanning technique, due to laser fluctuations where the change in the transmission is small compared to these fluctuations, Ménard et al. [9] proposed a method where the sample was mount on an oscillating-actuator in a Z-scan setup, thus, for an amplitude $S$ and frequency $F$ the transmission signal around (and near) position $z_0$ will be

$$T \cong T(z_0) + S\left(\frac{\partial T}{\partial z}\right)_{z=z_0} \sin(2\pi Ft), \quad (10)$$

where $t$ is time. Eq. (10) is valid as far as the oscillating amplitude $S$ is comparable to the Rayleigh range of the beam. If a lock-in amplifier is used with a reference signal $F$ coming from the piezoelectric actuator, then only the amplitude of the signal given by Eq. (11) is detected by the lock-in amplifier:

$$S\left(\frac{\partial T}{\partial z}\right)_{z=z_0}. \quad (11)$$

This background-free technique reduces the laser fluctuation noise and improves the sensitivity of the technique.

One feature of the EFTL is that its focal length can be modulated with different types of signal profiles (rectangular, triangular or sinusoidal) with a frequency range in its modulation between 0.2 up to 2000 Hz. This allowed us to modify the TF-scan into a DF-scan without using piezoelectric actuators, and also using the modulated signal as the reference for a lock-in amplifier; being the signal proportional to the derivative of the transmitted signal.

Thus, for the case of the DF-scan technique, $z$ is replaced by the focal distance $f$, and using Eq. (9) the derivative in Eq. (11) becomes

$$\left(\frac{\partial T}{\partial f}\right)_{f=f_0} = \frac{\left(f + \frac{d_s}{k_1^2 f^3}(f-d_s)\right)}{k_2}\sum_{m=0}^{N}\left\{\frac{2m[-B(f)]^{m+1}}{(m+1)}\prod_{n=0}^{m}\left[\frac{2(m-n)+\delta_{mn}}{2(m-n)+1}\right]\right\}, \quad (12)$$

where $k_1 = \frac{4\lambda C_f^2}{\pi D^2}$, and $k_2 = 2\ln(1+\sqrt{2})\frac{\beta(1-R)L_{eff}P_{avg}}{k_1\tau\nu\lambda}$.

*"Normalized" DF-scan*

For the case of TF-scan, normalization of the experimental data in order to fit it by using the analytical model is simply done by dividing the data with respect to the lock-in amplifier signal

$A$ for the shortest focal length. In a DF-scan setup this is not so direct because the signal detected by the lock-in amplifier at this same focal length is null. Knowing that for any focal length, programmed in a EFTL, the signal detected is

$$AS\left(\frac{\partial T}{\partial f}\right)_{f=f_0}, \quad (13)$$

in order to obtain the "normalized" DF-scan signal of Eq. (12), which is independent of the modulation parameters $A$ and $S$, the signal represented by Eq. (13) has to be divided by the factor $AS$, which is known in advance. To show this, in our experiment we programed the EFTL to vary its focal length using a square signal of frequency $F = 739$ Hz and 50% duty cycle, and for $S$ amplitudes corresponding to driven currents of 0.5 and 1.0 mA (see Fig. 6(left)). The voltage detected by the lock-in amplifier for the shortest focal length, when it is not being modulated, was 46.23 mV, corresponding to $A = 46.23/2$ mV. Division by two is because of the 50% duty cycle of the square signal. Fig. 6(right) shows the corresponding normalized DF-scan signals for both used amplitudes and for the corresponding derivative of the normalized TF-scan; they are in excellent agreement. These normalized DF-scan signals can know be fitted using Eq. (12).

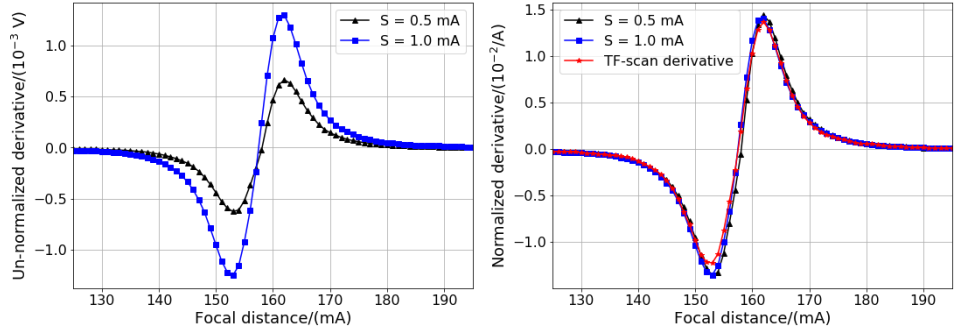

Fig. 6. (left) Un-normalized DF-scan signals for two oscillating amplitudes $S$, 0.5 mA and 1.0 mA. (right) Normalized DF-scan signals after dividing by amplitude $S$, voltage $A$ and 2 due to the 50% duty cycle. The corresponding Normalized TF-scan signal derivative is also shown.

## 4. Experimental data fitting protocol

A correct experimental data fitting is crucial in order to obtain a reliable value of the parameter of interest; in this case: the two-photon absorption coefficient. Is then desirable to know the values of all the experimental parameters with the highest precision possible in order to use only the parameter of interest as the fitting variable, especially if the technique requires the knowledge of a great number of parameters (for our case 10 experimental parameters have to be known). But in occasions this is not possible, and the experimental parameters are only known with a considerable uncertainty. As a first approach, one can use only the central values of the experimental parameters, ignoring the uncertainties, but this must probably will end in an inaccurate fitting and thus a wrong value of the parameter of interest (see for example Fig. 7(left) and set 1 in table 2). One naive approach will be to use more than one parameter as fitting variables, but because the model used does not have a unique solution, as is the case for TF-scan and DF-scan, one can end with a wrong value of the parameter of interest, although with a perfect curve fitting, as it is shown for example in Fig. 7(center).

Based on the assumption that the uncertainties of the experimental parameters correspond to random fluctuations that can be model with a Gaussian distribution, we implemented a protocol with the goal to obtain an interval that contains the real value of the parameter of interest, under a statistical approach. The protocol is the following:

1. For each experimental parameter pick a random value from the Gaussian distribution of the possible values.
2. Fit the experimental data an obtain the corresponding value for the parameter of interest.
3. Calculate a metric to evaluate the quality of the fit. If the metric satisfies a criteria keep the value of the parameter of interest. If not, discard it.
4. Repeat steps 1-3 until a distribution with a good sample size of acceptable values is obtained.
5. With the parameters of interest that were accepted calculate the average-weighted value, using each corresponding metric value as weights. Calculate the corresponding standard error. The average-weighted value corresponds to the parameter of interest.

To show the effectiveness of this approach a simulated experimental data "Real" from a sample with nonlinear TPA is presented. In table 1 the parameters used for the simulated data are presented.

**Table 1. Simulation: experimental parameters.**

| $L$ (mm) | $d_s$ (mm) | $\lambda$ (nm) | $D$ (mm) | $C_f$ |
|---|---|---|---|---|
| 0.8 | 116.0 | 790 | 2.0 | 1.36 |
| $P_{avg}$ (mW) | $\tau$ (fs) | $\nu$ (MHz) | $\alpha$ (1/m) | R |
| 145 | 71 | 90.9 | $2.64 \cdot 10^{-11}$ | 0.1567 |

To retrieve $\beta$ we supposed that the experimental parameters are known with a 10% uncertainty. Then, three sets of experimental parameters are pick from their normal distributions: set 1 and 2 correspond to a focal distance range from 8 to 16 cm with a total of 300 sample points, and set 3 correspond to the same range but 50 sample points. Finally, the data is fitted and $\beta$ is obtained using the central values "Direct" approach, our approach "Protocol", and the multiple fitting-parameters "Multiple" approach. For "Multiple" approach $D$, $\tau$, $P_{avg}$, and $d_s$ are also use as fitting parameters. In table 2 the results are presented.

**Table 2. Comparison of the $\beta$ values, in cm/GW, retrieved from the different data-fitting techniques.**

|  | Set 1 | | Set 2 | | Set 3 | |
|---|---|---|---|---|---|---|
|  | $\beta$ | Error % | $\beta$ | Error % | $\beta$ | Error % |
| Real | 3.4 |  | 3.4 |  | 3.4 |  |
| Direct | 2.2 | 36 | 3.8 | 13 | 5.3 | 53 |
| Multiple | 33.9 | 874 | 14.5 | 325 | 46.4 | 1265 |
| Protocol | 3.3 | 2 | 3.0 | 11 | 3.5 | 2 |

It is clear that the "Multiple" approach is not desirable because it most probably return a wrong $\beta$ value while giving an almost perfect data fitting curve, see Fig. 7(center), Fig. 8(center) and Fig. 9(center). The "Direct" approach depends directly on the experimental parameters chosen, and due to the uncertainty, this can mean that the value retrieved is close to the correct one (Fig. 8(left)) or far (Fig. 7(left) and Fig. 9(left)). Finally, our proposal, "Protocol", gives the certainty that it will always retrieve the value with the greatest possible accuracy (see Fig. 7(right), Fig. 8(right) and Fig. 9(right)).

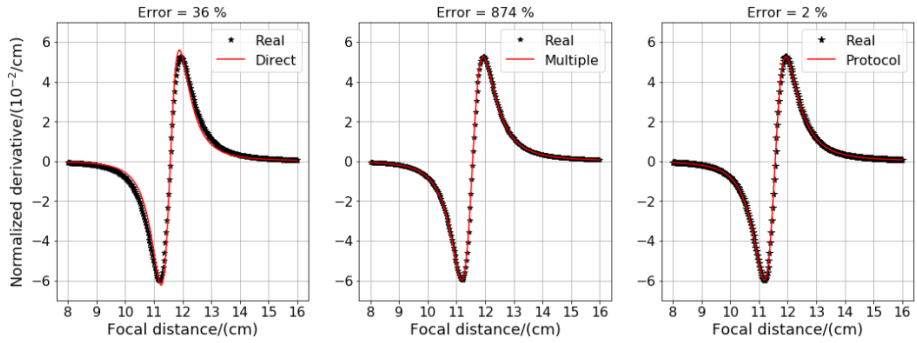

Fig. 7. Fitting results of the simulated data for Set 1 in table 2. (Left) "Direct" approach, (center) "Multiple" approach, (right) "Protocol" approach.

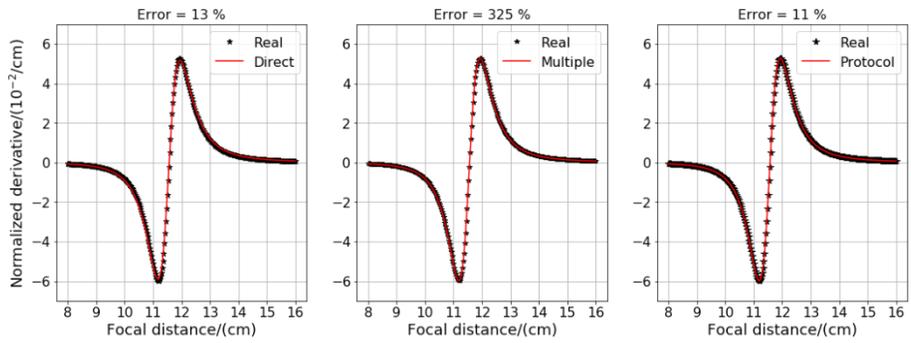

Fig. 8. Fitting results of the simulated data for Set 2 in table 2. (Left) "Direct" approach, (center) "Multiple" approach, (right) "Protocol" approach.

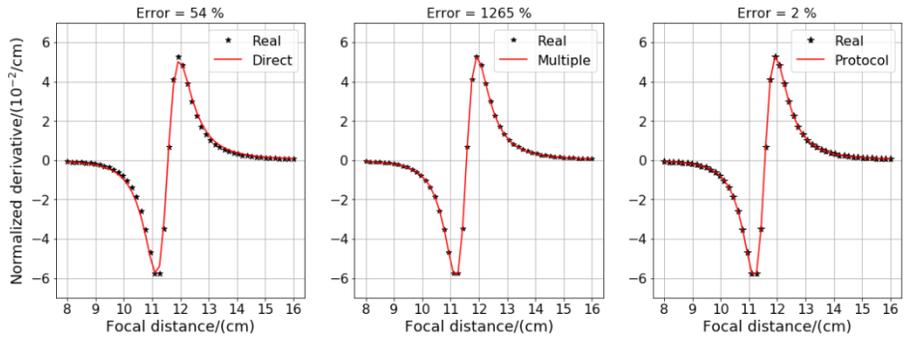

Fig. 9. Fitting results of the simulated data for Set 3 in table 2. (Left) "Direct" approach, (center) "Multiple" approach, (right) "Protocol" approach.

## 5. Experimental results

For the experimental implementation of the TF-scan and DF-scan we used a Ti:Sapphire oscillator laser with repetition rate of 90.9 MHz, pulse width of 71 fs, and laser emission centered at 790 nm. The average power at the entrance surface of the sample was 145 mW. The beam diameter at the EFTL was D = 2.0 mm, and was measured by a laser beam profiler. The EFTL is an OPTOTUNE-1030, controlled by an OPTOTUNE lens-driver that gives a maximum current of 300 mA with a resolution of 0.1 mA, delivering a focal length resolution of 0.017 mm. The laser Gaussian beam is focused at quasi-normal incidence in order to eliminate Fabry-Perot effects and multiple reflections. We used an integrating sphere with a large area Si-photodiode (PDA 50 THORLABS) to measure the transmitted laser light. This modification to the common setup compensates any lens-divergence and eliminates signal losses due to scattering from the sample-surface roughness. The current generated by the photodiode is sent to a STANFORD RESEARCH 830 dual channel Lock-in amplifier, controlled through a GPIB interface.

We have measured the TPA coefficient at 790 nm for ZnSe, CdS and CdSe. The experimental parameters for both techniques and all materials are listed in table 3, and the obtained TPA coefficients are listed in table 4. In Fig. 10 the experimental curves and curve fitting is presented for the case of CdSe.

Table 3. Experimental parameters for TF-scan and DF-scan, for CdS, ZnSe and CdSe.

| $\nu$ (MHz) | $d_s$ (mm) | $\lambda$ (nm) | $D$ (mm) | $C_f$ |
|---|---|---|---|---|
| 90.9 | 116.0 ± 0.5 | 790 ± 1 | 2.0 ± 0.8 | 1.36 ± 0.01 |
| $P_{avg}$ (mW) | $\tau$ (fs) | $L_{ZnSe}$ (mm) | $\alpha_{ZnSe}$ (1/m) | $R_{ZnSe}$ |
| 145 ± 5 | 71.0 ± 0.3 | 0.80 ± 0.01 | 4.772 | 0.182 ± 0.005 |
| $L_{CdS}$ (mm) | $\alpha_{CdS}$ (1/m) | $R_{CdS}$ | $L_{CdSe}$ (mm) | $\alpha_{CdSe}$ (1/m) |
| 0.85 ± 0.01 | $2.64 \cdot 10^{-11}$ | 0.157 ± 0.005 | 0.79 ± 0.01 | 369 ± 37 |
| $R_{CdSe}$ | | | | |
| 0.185 ± 0.005 | | | | |

Table 4. Comparison of $\beta$ values, in cm/GW at 790 nm, for ZnSe, CdS and CdSe.

| | ZnSe | | CdS | | CdSe | |
|---|---|---|---|---|---|---|
| | $\beta$ | Relative error % | $\beta$ | Relative error % | $\beta$ | Relative error % |
| DF-scan | 5.1 | 23 | 2.4 | 22 | 4.6 | 12 |
| TF-scan | 3.2 | 16 | 1.5 | 13 | 1.8 | 15 |
| Krauss [13] | 3.5[*] | >35 | 6.4[*] | >35 | | |

*at 780 nm.

From table 4 it is evident that there exists a discrepancy between TF-scan and DF-scan results, in particular, DF-scan always gives a bigger value. This discrepancy will be analyzed and explained in the fallowing subsection. For the value reported by Krauss et al. [13] for ZnSe, and from a statistical point of view, there is a probability of 82% and 35% that the difference is due to a random fluctuation with respect to TF-scan and DF-scan, respectively. For the case of CdS the probabilities are 3% and 8%, respectevely. A random deviation of this magnitude in the values is not consider rare. For CdSe we were not able to find a value for wavelengths close to 790 nm.

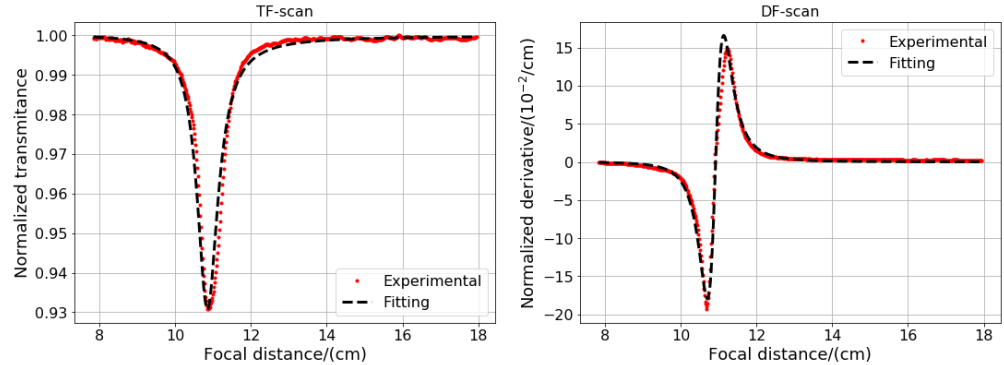

Fig. 10. Fitting result for CdSe. (Left) TF-scan, (right) DF-scan.

To compare our results with the values that have been reported, in a broader way, for the three materials, we used the expression derived by Wherrett [14] for TPA,

$$\beta \sim \frac{(2hc/(\lambda E_g) - 1)^{3/2}}{(2hc/(\lambda E_g))^5}, \quad (14)$$

where $E_g$ is the energy of the bandgap, $c$ is the speed of light in vacuum, and $h$ is the Planck's constant. In particular, the ratio $r_{790,\lambda}$ between the TPA at 790 nm with respect to the TPA value for the other wavelengths is given by

$$r_{790,\lambda} = \left(\frac{\lambda_{790}}{\lambda}\right)^{7/2} \left(\frac{2hc - \lambda_{790} E_g}{2hc - \lambda E_g}\right)^{3/2} \quad (15)$$

In Fig. 11 the ratio is plot for the three materials. Except for the case of CdSe, there is a probability of 5% or more that the differences are due to random fluctuations.

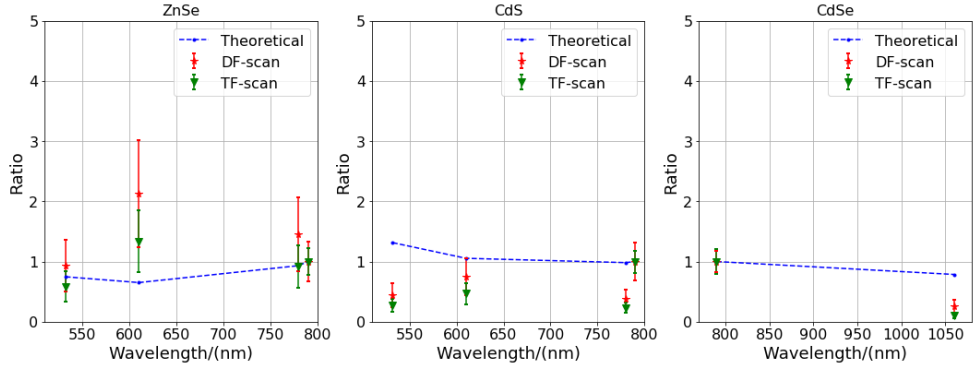

Fig. 11. Wherrett ratio. The values for the other wavelengths are taken from Krauss *et al.* [13] and V. Stryland *et al.* [13].

*Difference between TF-scan and DF-scan results*

We believe the discrepancy between the values obtained with DF-scan and TF-scan is due to fitting robustness directly related to curve shape. In our fitting criteria the metric minimizes the value of the sum of the distance between the calculated and experimental peaks of the curves. The existence of two peaks in DF-scan signals reduce the spectrum of possible fitting values, making the process more robust and accurate than with TF-scan signals. Then, the same result must be obtained for both, the derivative of TF-scan signal and the DF-scan signal. To validate this idea, first we perform a simulation with the values of table 1, and secondly we perform the fitting for the derivative of the TF-scan signal of CdSe and compare it with the results of table 4. Results are presented in table 5 and Fig. 12, showing, without doubt, that fitting the derivative of the signal of TF-scan or the signal of DF-scan outcomes a more accurate value.

Table 5. Comparison of the $\beta$ values, in cm/GW, retrieved from the simulated data of table 1.

|  | Simulated $\beta = 3.4$ | |
| --- | --- | --- |
|  | $\beta$ | Error % |
| TF-scan | 2.6 | 24 |
| DF-scan | 3.8 | 12 |
| TF-scan derivative | 3.6 | 6 |

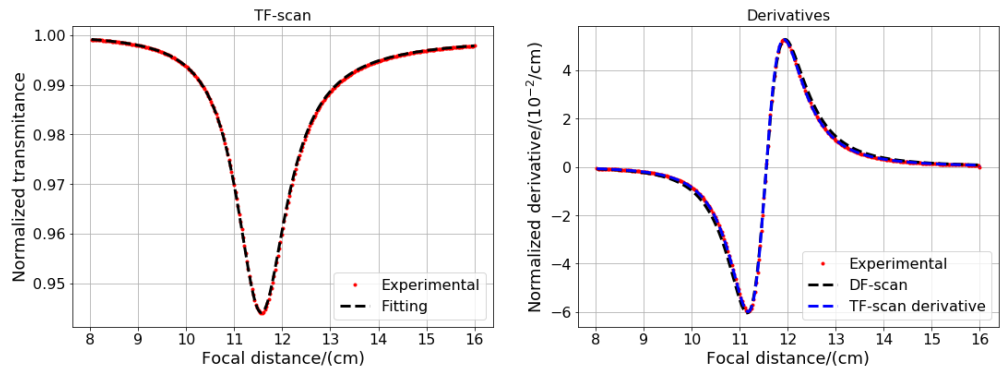

Fig. 11. Fitting curves of table 5 simulated-experimental data results. (Left) TF-scan, (right) DF-scan and TF-scan derivative.

## 6. Conclusions

DF-scan is a modification to the F-scan technique where the scanning is done over the rate of change of the transmission signal with respect of the focal distance of the EFTL, reducing drastically the sensibility to laser fluctuations, increasing the available laser power and simplifying the optical setup by eliminating the need of a chopper to modulate the signal. For the curve fitting step, where the experimental signal is fitted to an analytical model in order to obtain the TPA coefficient, a new protocol is proposed in order to secure the correct determination of the TPA parameter. The effectiveness of the proposal has been validated with respect to other curve fitting procedures. Finally, it was shown that it is more reliable to fit the derivatives of the experimental signals than the signal itself, either of the DF-scan signal or the derivative of the TF-scan signal.

### Acknowlegments

E. Rueda thanks Universidad de Antioquia for financial support. J. Serna acknowledges the support from Universidad Pontificia Bolivariana. H. Garcia and A. Hamad thanks Southern Illinois University, Edwardsville, for financial support.